\documentclass[10pt,twocolumn,amsmath,amssymb,aps,prl]{revtex4-2}

\begin{document}

\title{The Scale Invariance Behind Boltzmann Counting}

\author{M. H. Benetti}
\email{mhbenetti@usp.br}
\affiliation{
Departamento de Astronomia,
Instituto de Astronomia, Geof\'isica e Ci\^encias Atmosf\'ericas,
Universidade de S\~ao Paulo,
05508-090 S\~ao Paulo, SP, Brazil
}

\date{\today}

\begin{abstract}
Boltzmann counting possesses an exact scale invariance from which the specific entropy emerges without asymptotic approximation. Under thermodynamic replication, this invariance exposes the Gibbs excess at the multinomial level. Its resolution reveals a state-dependent scale structure whose additive refinement generates a continuum reference measure, giving a combinatorial origin to the reference structure of continuous statistical mechanics. The resulting entropy is relative to this measure, and for an ideal gas the scales acquire operational meaning through reversible work.

\end{abstract}

\maketitle
Phenomenologically, the Clausius relation at fixed particle number does
not determine how entropy scales with the amount of matter
\cite{Pauli1973,Jaynes1992}. Microscopically, Boltzmann counting yields
an extensive entropy when the number of accessible states is held fixed,
although this scaling is usually exposed only after an asymptotic
treatment of the factorials \cite{Planck1914}. The difficulty becomes
more delicate when the accessible state space itself grows. In the Gibbs
mixing problem, combining identical subsystems produces an entropy excess
that is simultaneously a manifestation of nonextensivity
\cite{Gibbs1902,Swendsen2018,Saunders2018,Sasa2022}. This excess is
usually suppressed by the conventional $N!$ correction, traditionally
associated with particle indistinguishability, although its classical
interpretation is not unique \cite{Swendsen2002}. In the continuum, the
problem takes a more general form: a coordinate-invariant entropy requires
a reference measure and is expressed as a relative entropy
\cite{Jaynes1963,Jaynes1968}. In specific dynamical models, the microscopic
physics can select the density of this measure \cite{MaynarTrizac2011},
but the continuous formalism itself does not determine whether its
reference structure has an origin in discrete counting. These issues
therefore concern two basic changes of description: increasing the amount
of matter and refining the accessible state space.

Here we show that the Boltzmann multinomial possesses a common-scale
invariance that connects these two descriptions. Once made explicit, this
scale places the specific entropy in an exact finite-$N$ Riemann-sum form,
without asymptotic factorials. Under thermodynamic replication, the Gibbs
excess appears already at the multinomial level, before any mixing process
is introduced, and can be traced to the cardinality of the replicated
system rather than to microscopic particle distinguishability. Separating
the common scale into global and local components removes this excess. The local
components survive as state-dependent scale ratios, so the entropy becomes
relative to them and equal-energy states need not carry equal populations.
Refinement imposes an additive composition law on these ratios, from which
the continuum reference measure follows, together with a natural
interpretation of its relative weights as statistical accessibility. In a homogeneous ideal gas, the same
structure accounts for extensivity and the vanishing entropy of mixing of
identical gases. More generally, statistical accessibility can retain
microscopic information about physical distinctions and transmit it to
free energy and reversible work, linking microscopic statistical structure
to macroscopic thermodynamics.

\noindent{\textit{Scale-invariant Boltzmann counting.}}
Consider $N$ particles distributed among $K$ states, with
$\sum_i n_i=N$. The usual Boltzmann multiplicity is written in compact
factorial form \cite{Boltzmann1877}, where no microscopic scale appears
explicitly. When its factorials are expanded into elementary chains, a
remarkable invariance emerges:
\begin{equation}
W
=
\frac{N!}{\prod_i n_i!}
=
\frac{\displaystyle\prod_{j=1}^{N}j}
{\displaystyle\prod_i^K\prod_{j=1}^{n_i}j}
=
\frac{\displaystyle\prod_{j=1}^{N}j\Delta_N}
{\displaystyle\prod_i^K\prod_{j=1}^{n_i}j\Delta_N}.
\label{eq:WBproduct}
\end{equation}
The last equality holds for any $\Delta_N>0$, because
\begin{equation}
W[\Delta_N]
=
(\Delta_N)^{N-\sum_i n_i}W
=
W .
\label{eq:CommonScaleInvariance}
\end{equation}
Boltzmann's multinomial is therefore homogeneous of degree zero under
uniform rescaling. Among the equivalent representations, take the
complete chain to have unit length, $N\Delta_N=1$, so that
$\Delta_N=1/N$. Growing endpoints $N$ and $n_i$ are replaced by fixed
normalized endpoints $1$ and $p_i=n_i/N$. The specific entropy becomes
an exact Riemann sum:
\begin{equation}
\frac{S}{Nk_B}
=
\frac{S\Delta_N}{k_B}
=
\sum_{j=1}^{N}\ln(j\Delta_N)\Delta_N
-
\sum_i\sum_{j=1}^{n_i}\ln(j\Delta_N)\Delta_N .
\label{eq:FiniteRiemann}
\end{equation}
For fixed $K$, consider the thermodynamic limit $N\to\infty$ along a
sequence satisfying $n_i/N\to p_i>0$ and $\Delta_N\to0^+$. Factorial
chains approach $(0,1]$ and $(0,p_i]$, yielding
\begin{equation}
\lim_{N\rightarrow\infty}\frac{S}{Nk_B}
=
\int_{0^+}^{1}\ln x\,dx
-
\sum_i\int_{0^+}^{p_i}\ln x\,dx
\equiv s .
\label{eq:ShannonLimit}
\end{equation}
Evaluation of these integrals gives
$s=-\sum_i p_i\ln p_i$ and hence $S=Nk_Bs+o(N)$, with the subextensive
term reflecting the finite resolution of the Riemann sums. No asymptotic
approximation of the factorials is required, and the extensive factor is
already explicit before the limit. As $N$ increases,
$\Delta_N=1/N\to0^+$ turns the discrete counting chains into continuous
intervals. Note that, in this limit, all occupation  information is captured by the
upper endpoint $p_i$.

\noindent{\textit{Thermodynamic replication.}}
The preceding section established extensivity when the particle number
grows at fixed $K$. Increasing the amount of matter, however, does not
uniquely determine how the accessible state space changes. For an
enlargement by a factor $\lambda$, one may keep the original set of
states fixed, so that $n_i\to\lambda n_i$. Alternatively, the microscopic
state structure can itself be replicated, so that $N\to\lambda N$ and $K\to\lambda K$, with copied states $(i,a)$ satisfying $n_{ia}=n_i$, $a=1,\ldots,\lambda$. These possibilities lead to
\begin{equation}
\underbrace{
\frac{
\displaystyle\prod_{j=1}^{\lambda N}
j\,\frac{\Delta_N}{\lambda}
}{
\displaystyle\prod_{i=1}^{K}
\prod_{j=1}^{\lambda n_i}
j\,\frac{\Delta_N}{\lambda}
}
}_{\text{fixed-$K$ growth}}
\qquad
\underbrace{
\frac{
\displaystyle\prod_{j=1}^{\lambda N}
j\,\frac{\Delta_N}{\lambda}
}{
\displaystyle\prod_{a=1}^{\lambda}
\prod_{i=1}^{K}
\prod_{j=1}^{n_i}
j\,\frac{\Delta_N}{\lambda}
}
}_{\text{replication}} .
\label{eq:ReplicationScales}
\end{equation}
By the invariance~\eqref{eq:CommonScaleInvariance}, both expressions are
Boltzmann multinomials with the resolution
$\Delta_N/\lambda=1/(\lambda N)$. In the first case, the endpoint
$(\lambda n_i)\Delta_N/\lambda=p_i$ is invariant, preserving the
specific entropy and the thermodynamic extensivity derived above.

However, the more general test is the
replication of the macroscopic state. In the second expression, the common scale reveals a
different behavior: copied states retain the occupation $n_i$, so the
endpoint becomes $n_i\Delta_N/\lambda=p_i/\lambda$ and is no longer
invariant. The multinomial treats the $\lambda$ copies of state $i$ as
separate counting sectors, replacing the single fraction $p_i$ by
$\lambda$ counted contributions $p_i/\lambda$. This additional
resolution introduces a multiplicity associated with the cardinality
$\lambda$. In the thermodynamic limit, this multiplicity contributes
$k_B\ln\lambda$ per particle, or $\lambda Nk_B\ln\lambda$ to the total
entropy. For $\lambda=2$, we have $2Nk_B\ln2$. That is, the same
excess term obtained by Gibbs is recovered without a
mixing process between identical gases or invoking  particle distinguishability.   

The standard correction is the $N!$ factor. The present
construction suggests a different route. Once the excess is traced to the
common scale, the natural step is to separate
the global from the local component. A copied occupation chain need not carry information about how many replicas of the whole system are present. Thus

\begin{equation}
\underbrace{
\frac{
\displaystyle\prod_{j=1}^{\lambda N}
j\,\frac{\Delta_N}{\lambda}
}{
\displaystyle\prod_{a=1}^{\lambda}
\prod_{i=1}^{K}
\prod_{j=1}^{n_i}
j\,\Delta_i
}
}_{\text{local-scale replication}},
\label{eq:LocalScaleReplication}
\end{equation}
where $\Delta_i$ defines this local quantity, leaving the endpoint
$n_i\Delta_i$ invariant. The same
separation can be written for the original counting as
\begin{equation}
W[\Delta_N,\{\Delta_i\}]
=
\frac{\displaystyle\prod_{j=1}^{N}j\Delta_N}
{\displaystyle\prod_i\prod_{j=1}^{n_i}j\Delta_i}
=
W\prod_i
\left(\frac{\Delta_N}{\Delta_i}\right)^{n_i}.
\label{eq:LocalCounting}
\end{equation}
Here $W\equiv W[\Delta_N,\{\Delta_N\}]$.
For unequal scales, $W[\Delta_N,\{\Delta_i\}]$ is positive,
with the ratio $\Delta_N/\Delta_i$ providing a relative statistical weight for
state $i$, which need not be a normalized probability or an integer
multiplicity.  This freedom sets Eq. (\ref{eq:LocalCounting}) apart  
from the probability multinomial,
$P_{\rm mult}=W\prod_i q_i^{n_i}$, associated with the Kullback-Leibler relative entropy, where 
$\sum_i q_i=1$ \cite{KullbackLeibler1951}, and from the discrete degeneracies of the
Maxwell--Boltzmann form,
$W_{\rm MB}=W\prod_i g_i^{n_i}$, where $g_i\in\mathbb{N}$
\cite{Tolman1938}.

\noindent{\textit{Entropy and equilibrium.}}
We first examine the consequences of this separation at fixed $K$. For
$p_i=n_i/N=n_i\Delta_N$, introduce
\begin{equation}
f_i\equiv n_i\Delta_i,
\qquad
p_i=\frac{\Delta_N}{\Delta_i}f_i.
\label{eq:LocalOccupation}
\end{equation}
Here $f_i$ measures the occupation of state $i$ in its own local
coordinate and specifies the respective endpoint. When
$\Delta_i=\Delta_N$, one recovers $f_i=p_i$. Using
Eq.~\eqref{eq:LocalCounting}, the entropy per particle is
\begin{equation}
\frac{S}{Nk_B}
=
\sum_{j=1}^{N}\ln(j\Delta_N)\Delta_N
-
\sum_i
\frac{\Delta_N}{\Delta_i}
\sum_{j=1}^{n_i}\ln(j\Delta_i)\Delta_i .
\label{eq:LocalFiniteRiemann}
\end{equation}
In the thermodynamic limit, $\Delta_N,\Delta_i\to0^+$ while $p_i$ and
$f_i$ remain finite. The
counting chains approach the intervals $(0,1]$ and $(0,f_i]$,
respectively. Hence Eq.~\eqref{eq:LocalFiniteRiemann} becomes
\begin{equation}
\lim_{N\rightarrow\infty}\frac{S}{Nk_B}
=
\int_{0^+}^{1}\ln x\,dx
-
\sum_i\frac{p_i}{f_i}
\int_{0^+}^{f_i}\ln x\,dx
\equiv s .
\label{eq:LocalIntegralEntropy}
\end{equation}
The normalization
$\sum_i(\Delta_N/\Delta_i)f_i=\sum_i p_i=1$, together with
Eq.~\eqref{eq:LocalOccupation}, reduces the limiting integrals at
leading thermodynamic order to
\begin{equation}
S
=
-Nk_B\sum_i p_i
\ln\left[
\frac{p_i}{\Delta_N/\Delta_i}
\right],
\label{eq:LocalEntropy}
\end{equation}
which has the form of a discrete relative entropy, with
$\Delta_N/\Delta_i$ playing the role of state-dependent reference
weights \cite{Niven2009}.

From Eq.~\eqref{eq:LocalEntropy}, the equilibrium distribution follows
by maximizing $S=Nk_Bs$ under normalization and specific-energy
constraints. With Lagrange multipliers $\alpha$ and $\beta$, one obtains
\begin{equation}
f_i^{\rm eq}
=
e^{-\alpha-\beta\epsilon_i},
\qquad
p_i^{\rm eq}
=
\frac{\Delta_N}{\Delta_i}
e^{-\alpha-\beta\epsilon_i}.
\label{eq:LocalEquilibriumProbability}
\end{equation}
Normalization fixes
$e^\alpha=\sum_i(\Delta_N/\Delta_i)e^{-\beta\epsilon_i}$.
For two states $i_1$ and $i_2$ with
$\epsilon_{i_1}=\epsilon_{i_2}$,
Eq.~\eqref{eq:LocalEquilibriumProbability} gives $f_{i_1}^{\rm eq}=f_{i_2}^{\rm eq}$, but their probabilities satisfy
\begin{equation}
\frac{p_{i_1}^{\rm eq}}{p_{i_2}^{\rm eq}}
=
\frac{\Delta_{i_2}}{\Delta_{i_1}}.
\label{eq:EqualEnergyCapacity}
\end{equation}
Equal energy does not imply equal populations; their ratio is set by
the local counting scales.

\noindent{\textit{The reference measure.}}
The remaining question is how the relative factors
$\Delta_N/\Delta_i$ compose when the same accessible structure is
represented as a single state or subdivided into smaller parts. Let the
occupation of state $i$ be split into suboccupations $n_{i\alpha}$,
with local scales $\Delta_{i\alpha}$, such that
$n_i=\sum_\alpha n_{i\alpha}$. The counting of state $i$ is unchanged
by this refinement, so one sums over all allowed sets
$\{n_{i\alpha}\}$ compatible with the given $n_i$. Inserting these
terms into Eq.~\eqref{eq:LocalCounting} and applying the multinomial
theorem yields
\begin{equation}
\left(
\Delta_i\sum_\alpha\frac{1}{\Delta_{i\alpha}}
\right)^{n_i}
=
1
\quad\Longrightarrow\quad
\frac{\Delta_N}{\Delta_i}
=
\sum_\alpha
\frac{\Delta_N}{\Delta_{i\alpha}} .
\label{eq:RelativeScaleAdditivity}
\end{equation}
The scale ratio $\Delta_N/\Delta_i$ obeys an additive composition law
under subdivision. For $m$ equivalent subparts, we have
$p_{i\alpha}=p_i/m$ and
$\Delta_N/\Delta_{i\alpha}=(\Delta_N/\Delta_i)/m$.
Remarkably, replication produces an analogous transformation: each copy
carries $p_{ia}=p_i/\lambda$ and
$(\Delta_N/\Delta_i)/\lambda$. In both cases, the particle fraction and
scale ratio are repartitioned by the same factor, so that
$f_{i\alpha}=f_i$ and $f_{ia}=f_i$.
For a given $f_i$, a larger $\Delta_N/\Delta_i$ corresponds to a larger
fraction $p_i$ of the total population, supporting its interpretation as
the relative statistical accessibility of state $i$. Define this quantity as
\begin{equation}
A_i
\equiv
\frac{\Delta_N}{\Delta_i},
\qquad
p_i= A_i f_i,
\qquad
\sum_a A_{ia}
=
\sum_\alpha A_{i\alpha}
=
 A_i.
\label{eq:DiscreteAccessibility}
\end{equation}
Physically distinct operations repartition the relative accessibility
without changing its total value. As the partition is successively
refined and $\max_i A_i\to0$, this additivity generates a
measure $ A$ over the one-particle state space
$\Gamma_1$,  while the particle fractions define the probability measure
$P$. In this limit, the discrete relation becomes
\begin{equation}
f_i
=
\frac{p_i}{ A_i}
\longrightarrow
\frac{dP}{d A},
\qquad
s_{ A}[P]
=
-\int_{\Gamma_1}
\ln\left(\frac{dP}{d A}\right)dP .
\label{eq:RelativeContinuum}
\end{equation}
The quantity $dP/dA$ represents probability per unit
statistical accessibility. Coordinate-invariant continuous entropy
requires probability to be expressed relative to a reference measure
\cite{Jaynes1963,Jaynes1968}. A closely related realization of this reference structure appears in the
measure problem for continuous collision processes, where the microscopic
collision law determines a nonuniform sampling density
\(m_X(x)\propto\Lambda_X^{-1}(x)\) \cite{MaynarTrizac2011}. Writing
\(dA=m_X(x)\,dx\), Eq.~\eqref{eq:RelativeContinuum} takes precisely the
measure-corrected form introduced to avoid improper phase-space weighting in
continuous descriptions, with
\(dA= m_X(x)\,dx\propto\Lambda_X^{-1}(x)\,dx\).
In such collision models, the specific density of the measure is selected by
the microscopic dynamics through the Jacobian of the collision law. Here,
instead, the structure of the reference measure and its additive composition
law emerge from Boltzmann counting, while its particular microscopic density
must still be supplied by the underlying physics.
Finally, the discrete equilibrium structure has a  continuum counterpart.
From Eq.~\eqref{eq:RelativeContinuum},
\begin{equation}
\frac{dP_{\rm eq}}{d A}
=
e^{-\alpha-\beta\epsilon(\sigma)},
\qquad
\sigma\in\Gamma_1.
\label{eq:ContinuumEquilibriumMeasure}
\end{equation}
Energy therefore fixes the equilibrium probability per unit statistical
accessibility, while $dA$ determines the statistical weight of
each region of state space. Equal-energy regions can consequently carry
different total probabilities.

The construction therefore connects the two changes of description.
Replication fixes how the relative weights transform with the amount of
matter, and refinement fixes the additive composition of the reference
measure. In this sense, Eq.~\eqref{eq:LocalCounting}
encodes the relative structure and transformation law of the
continuum measure \cite{Jaynes1963,Jaynes1968}, but not its microscopic
basis. The latter must arise from the underlying  physics.

\noindent{\textit{Thermodynamic interpretation.}}
For a fine partition of the one-particle phase space, let
$\Delta\Gamma_i$ denote the Liouville volume of cell $i$. Since
$ A_i=\Delta_N/\Delta_i$ and $\Delta_N=1/N$, the relative accessibility
density is
\begin{equation}
\rho_{ A,i}
=
\frac{ A_i}{\Delta\Gamma_i}
=
\frac{1}{N\Delta_i\Delta\Gamma_i}.
\end{equation}
For the conventional homogeneous monatomic ideal gas, equal Liouville phase-space
volumes carry equal statistical accessibility, so $\rho_{A,i}$ is
uniform. Under an equivalent subdivision, the refinement law
Eq.~\eqref{eq:RelativeScaleAdditivity} partitions $A_i$ and
$\Delta\Gamma_i$ in the same proportion, leaving their ratio independent
of resolution. Denoting this common ratio by $A/\Delta\Gamma$, the continuum
accessibility measure for a system with $N$ particles becomes
\begin{equation}
d A_N
=
\frac{ A}{\Delta\Gamma}
\,d^3r\,d^3p .
\label{eq:HomogeneousMeasure}
\end{equation}
Equation~\eqref{eq:HomogeneousMeasure}, with
$\epsilon=p^2/(2m)$, yields
\begin{equation}
\frac{dP_{\rm eq}}{d A_N}
=
e^{-\alpha-\frac{\beta p^2}{2m}},
\qquad
dP_{\rm eq}
=
\frac{ A}{\Delta\Gamma}
e^{-\alpha-\frac{\beta p^2}{2m}}
\,d^3r\,d^3p .
\label{eq:IdealGasDistribution}
\end{equation}
Normalization gives the constant $\alpha$, the mean energy satisfies
$u=U/N=3/(2\beta)$. In this case
$s_{\rm eq}=S_{\rm eq}/Nk_B=\alpha+\beta u$, and the equilibrium
specific entropy, at leading order, is
\begin{equation}
\frac{S_{\rm eq}}{Nk_B}
=
\alpha+\frac{3}{2}
=
\ln\left[
V\frac{ A}{\Delta\Gamma}
\left(
\frac{U}{N}\frac{4\pi e m}{3}
\right)^{3/2}
\right].
\label{eq:IdealGas}
\end{equation}
The dependence required for extensivity is already contained in the
accessibility measure. Under replication, $N\to\lambda N$,
$V\to\lambda V$, and $U\to\lambda U$. The discrete transformation
$A_{ia}=A_i/\lambda$ implies, at the same local Liouville resolution,
$A/\Delta\Gamma
\longrightarrow
A/(\lambda\Delta\Gamma)$.
Consequently, $VA/\Delta\Gamma$ and $U/N$ remain invariant, and
$S_{\rm eq}\to\lambda S_{\rm eq}$.
The usual thermodynamic derivatives give
$\beta=(k_BT)^{-1}$, $U=3Nk_BT/2$, and $PV=Nk_BT$. Any dependence of accessibility 
on $U$ or $V$ may contribute to the caloric relation and the
equation of state.  For the Gibbs problem, removing an inert partition between two
identical homogeneous gases cannot alter their local statistical structure;
otherwise, the operation itself would create a physical distinction
between them. With the local structure preserved, the 
scaling above applies, and Eq.~\eqref{eq:IdealGas} gives
$\Delta S_{\rm mix}=0$.

The question now concerns how the information associated with relative
accessibility enters reversible work. Equilibrium states may differ in volume, accessibility, or both,
although the counting does not specify the microscopic origin of the
accessibility difference. To separate these thermodynamic contributions,
consider a reversible isothermal transformation from an initial
equilibrium state I to a final equilibrium state F, at fixed $N$ and $U$
and with the same phase-space resolution. Equation~\eqref{eq:IdealGas}
then implies
\begin{equation}
W_{\max}
=
T\Delta S_{\rm eq}
=
Nk_BT
\left[
\ln\frac{V_{\rm F}}{V_{\rm I}}
+
\ln\frac{ A_{\rm F}}{ A_{\rm I}}
\right].
\label{eq:MacroscopicAccessibilityWork}
\end{equation}
Such information becomes thermodynamically operative when an interaction
can act selectively on the microscopic property it reflects
\cite{Jaynes1992,Yadin2021}. A semipermeable membrane provides the
standard example: selectivity allows a component to expand reversibly
into a larger spatial domain, converting the microscopic information into
work through the volume term in
Eq.~\eqref{eq:MacroscopicAccessibilityWork}. At fixed volume, if such equilibrium states can be connected by a
reversible isothermal transformation, Eq.~\eqref{eq:MacroscopicAccessibilityWork}
reduces to
\begin{equation}
W_{\max}
=
Nk_BT
\ln\frac{A_{\rm F}}{A_{\rm I}} .
\label{eq:AccessibilityWorkIdealGas}
\end{equation}

Finally, inverting the thermodynamic connection provides an operational
probe of the local scales. Conditioning the equilibrium measure on a cell $i$ assigns the
internal entropy
$s_A[P_i]=\ln A_i=\ln(\Delta_N/\Delta_i)$. This quantity plays the role
of an internal entropy of the coarse-grained state, consistent with
stochastic thermodynamics \cite{Esposito2012}. For two equal-energy cells
$\epsilon_{i_1}=\epsilon_{i_2}$, their free-energy difference is therefore
purely entropic and is accessible through reversible isothermal work
\cite{Parrondo2015}:
\begin{equation}
\frac{ A_{i_2}}{ A_{i_1}}
=
\frac{\Delta_{i_1}}{\Delta_{i_2}}
=
\exp\left[
\frac{w_{i_1}^{\max}-w_{i_2}^{\max}}{k_BT}
\right].
\label{eq:OperationalAccessibility}
\end{equation}
\noindent{\textit{Conclusion.}}
The present work brings to light a structure in Boltzmann microstate counting
that, to our knowledge, has not previously been explored in this form. When
the factorials are unfolded into elementary counting chains, a striking
invariance emerges, leading directly to the specific entropy in its occupation
fraction representation. Following this scale as the amount of matter
increases then reveals an unexpected perspective on both thermodynamic
extensivity and the Gibbs excess, and points naturally to where the counting
must be reorganized: by separating the global scale from the local resolution
of the occupation chains.

Remarkably, the same elementary structure gives relative entropy a natural
combinatorial footing. From this single construction emerge state-dependent
equilibrium weights, the additive law under refinement, the continuum
reference measure, and the vanishing mixing entropy of identical gases. The
local scales thus appear as part of the statistical constitution of the
description itself, supplying the relative structure from which the reference
measure is built, while its specific microscopic realization is fixed by the
underlying physics. When physically resolvable, these relative scales become
thermodynamically operative through reversible work.

\begin{acknowledgments}
M.H.B. is supported by FAPESP/CNPq (24/14163-3).
\end{acknowledgments}


\begin{thebibliography}{99}

\bibitem{Pauli1973}
W. Pauli,
\textit{Thermodynamics and the Kinetic Theory of Gases},
Pauli Lectures on Physics, Vol.~3, edited by C. P. Enz
(MIT Press, Cambridge, MA, 1973).

\bibitem{Jaynes1992}
E. T. Jaynes,
in \textit{Maximum Entropy and Bayesian Methods: Seattle, 1991},
edited by C. R. Smith, G. J. Erickson, and P. O. Neudorfer
(Kluwer Academic, Dordrecht, 1992), pp.~1--21.

\bibitem{Planck1914}
M. Planck,
\textit{The Theory of Heat Radiation},
2nd ed., translated by M. Masius
(P. Blakiston's Son \& Co., Philadelphia, 1914).


\bibitem{Gibbs1902}
J. W. Gibbs,
\textit{Elementary Principles in Statistical Mechanics}
(Yale University Press, New Haven, 1902).

\bibitem{Swendsen2018}
R. H. Swendsen,
Entropy \textbf{20}, 450 (2018).

\bibitem{Saunders2018}
S. Saunders,
\textit{Entropy} \textbf{20}, 552 (2018).

\bibitem{Sasa2022}
S.-i. Sasa, K. Hiura, N. Nakagawa, and A. Yoshida,
J. Stat. Phys. \textbf{189}, 31 (2022).

\bibitem{Swendsen2002}
R. H. Swendsen,
Statistical Mechanics of Classical Systems with Distinguishable Particles,
\textit{J. Stat. Phys.} \textbf{107}, 1143--1166 (2002).


\bibitem{Jaynes1963}
E. T. Jaynes,
in \textit{Statistical Physics},
1962 Brandeis Lectures in Theoretical Physics, Vol.~3,
edited by K. W. Ford
(W. A. Benjamin, New York, 1963), pp.~181--218.

\bibitem{Jaynes1968}
E. T. Jaynes,
IEEE Trans. Syst. Sci. Cybern. \textbf{4}, 227 (1968).

\bibitem{Boltzmann1877}
L. Boltzmann,
Wiener Berichte \textbf{76}, 373 (1877).



\bibitem{KullbackLeibler1951}
S. Kullback and R. A. Leibler,
Ann. Math. Stat. \textbf{22}, 79 (1951).

\bibitem{Tolman1938}
R. C. Tolman,
\textit{The Principles of Statistical Mechanics}
(Oxford University Press, Oxford, 1938).

\bibitem{Niven2009}
R. K. Niven,
Eur. Phys. J. B \textbf{70}, 49 (2009).



\bibitem{MaynarTrizac2011}
P. Maynar and E. Trizac,
Phys. Rev. Lett. \textbf{106}, 160603 (2011).

\bibitem{Yadin2021}
B. Yadin, B. Morris, and G. Adesso,
\textit{Nat. Commun.} \textbf{12}, 1471 (2021).

\bibitem{Esposito2012}
M. Esposito,
Phys. Rev. E \textbf{85}, 041125 (2012).

\bibitem{Parrondo2015}
J. M. R. Parrondo, J. M. Horowitz, and T. Sagawa,
Nat. Phys. \textbf{11}, 131 (2015).

\end{thebibliography}
\end{document}